\documentclass[10pt,letterpaper]{article}

\usepackage{graphicx}

\usepackage[margin=2cm]{geometry}

\usepackage{amsmath,amssymb}
\usepackage{bbm}

\usepackage{changepage}

\usepackage[utf8x]{inputenc}

\usepackage{textcomp,marvosym}

\usepackage{cite}

\usepackage{nameref,hyperref}

\usepackage[right]{lineno}

\usepackage{microtype}
\DisableLigatures[f]{encoding = *, family = * }

\usepackage[table]{xcolor}

\usepackage{array}

\newcolumntype{+}{!{\vrule width 2pt}}

\newlength\savedwidth

\makeatletter
\renewcommand{\@biblabel}[1]{\quad#1.}
\makeatother

\usepackage{placeins}

\usepackage{mdframed,threeparttable}

\def \draft {1}

\usepackage{xparse}
\usepackage{ifthen}
\DeclareDocumentCommand{\comment}{m o o o o}
{\ifthenelse{\draft=1}{
    \textcolor{red}{\textbf{C : }#1}
    \IfValueT{#2}{\textcolor{blue}{\textbf{A1 : }#2}}
    \IfValueT{#3}{\textcolor{ForestGreen}{\textbf{A2 : }#3}}
    \IfValueT{#4}{\textcolor{red!50!blue}{\textbf{A3 : }#4}}
    \IfValueT{#5}{\textcolor{Aquamarine}{\textbf{A4 : }#5}}
 }{}
}
\newcommand{\todo}[1]{
\ifthenelse{\draft=1}{\textcolor{red!50!blue}{\textbf{TODO : \textit{#1}}}}{}
}

\usepackage{ragged2e}

\title{Modeling growth of urban firm networks} 

\date{}

\author{
Juste Raimbault\textsuperscript{1,2,3*},
Natalia Zdanowska\textsuperscript{1,3},
Elsa Arcaute\textsuperscript{3}\bigskip\\
\textsuperscript{1} Centre for Advanced Spatial Analysis, UCL, London, United Kingdom
\\
\textsuperscript{2} UPS CNRS 3611 ISC-PIF, Paris, France
\\
\textsuperscript{3} UMR CNRS 8504 G{\'e}ographie-cit{\'e}s, Paris, France\bigskip\\
* \texttt{juste.raimbault@polytechnique.edu}
}

\begin{document}

\maketitle

\begin{abstract}
The emergence of interconnected urban networks is a crucial feature of globalisation processes. Understanding the drivers behind the growth of such networks -- in particular urban firm networks --, is essential for the economic resilience of urban systems. We introduce in this paper a generative network model for firm networks at the urban area level including several complementary processes: the economic size of urban areas at origin and destination, industrial sector proximity between firms, the strength of links from the past, as well as the geographical and socio-cultural distance. An empirical network analysis on European firm ownership data confirms the relevance of each of these factors. We then simulate network growth for synthetic systems of cities, unveiling stylized facts such as a transition from a local to a global regime or a maximal integration achieved at an intermediate interaction range. We calibrate the model on the European network, outperforming statistical models and showing a strong role of path-dependency. Potential applications of the model include the study of mitigation policies to deal with exogenous shocks such as economic crisis or potential lockdowns of countries, which we illustrate with an application on stylized scenarios.
\end{abstract}

\justify

% Keywords
% Urban firm networks; Simulation model; Model calibration

\section*{Introduction}

Networks constitute the new social morphology of our societies, since the intensification of globalisation of the economy in the late 20th century \cite{castells2000networksociety}. A world-city system has since then emerged \cite{taylor2001specification}, characterized by highly interconnected urban centres at the global level \cite{sassen1991}. These interactions can be of different nature and determined by economic, socio-cultural, or geopolitical proximity inducing spatial and non-spatial patterns \cite{martinus2018global}. Interactions of economic nature between transnational firms \cite{derudder2018central}, are together with international trade patterns among the most representative of the current economic trends \cite{taylor2001specification}. Understanding the drivers of growth and geographical properties of such an urban network, can be used to foster innovation in specific cities and to shape public policies for local industrial clusters \cite{turkina2016structure}. In addition, straightening these city-interactions can be crucial for revitalising certain geographical areas \cite{Clarke2018}, by developing strategic industries for global integration of regions \cite{dawley2019creating}, cities \cite{gluckler2016relational} and countries \cite{martinus2019brokerage}. Exploring city/firm interactions \cite{storme2019introducing} and the position of cities within these networks \cite{gluckler2016relational} can also permit to infer the consequences of future exogenous shocks such as an economic crisis, a post-Brexit scenario for the European Union or a single market collapse in the post-COVID19 future.

The use of network models to understand processes underlying the emergence and dynamics of such networks has already been tackled in the literature, but still not extensively and never explicitly at city level \cite{taylor2001specification}. In fact generative urban network models combining geographical factors with network topological properties exist at the regional scale {\cite{dai2016simulating}}. Trade networks have been extensively studied from the complex networks perspective, but these are generally driven at the country-level \cite{garlaschelli2005structure} because of a lack of city-level data. Input-output models, mostly used in regional science to represent flows between geographical areas, are considered in some cases as a type of urban generative model \cite{jin1993generation}. In this stream of research, spatial interaction models \cite{dennett2013multilevel} can form the basis of urban network models for different types of flows \cite{dai2016generative} or physical infrastructure \cite{raimbault2018indirect}.

In this paper we choose to focus on urban networks of firms to question and capture geographical and economic processes as internationalisation, metropolisation, regionalisation and specialisation of cities within a generative model. The specific question we tackle is the complementarity of different processes to generate such urban networks. The originality of the present paper is to examine these issues within the system of cities framework \cite{berry1964cities}. In this approach the position and dynamics of cities in the socio-economic world-wide space can be considered by their interactions with other cities \cite{pumain2018evolutionary}. We examine the interactions of European cities within firm linkages defined by corporate ownership links. Since transnational firms structure is one of the determinant of the global economic space, it provides one accurate proxy to unveil geo-economic structures \cite{2019arXiv191014652Z}. 

More precisely, our contribution consists in the following: (i) we propose an empirical analysis of the firm ownership urban network in the European Union, based on the Bureau Van Dijk's \emph {AMADEUS} database \cite{AMADEUS2018}; (ii) we introduce a generative network model to simulate the growth of such linkages at urban areas scale, which combines multiple factors influencing link formation, such as the economic intensity of the urban areas corresponding to the origin and destination nodes, the industrial sector proximity, the strength of previous links, and the geographical and socio-cultural distance; (iii) the model allows us to compare the effect of different factors on the final network structure, which is extensively studied through model sensitivity analysis and exploration; and (iv) we calibrate the model on the empirical network.

The rest of this paper is organized as follows: we firstly describe the generative model for urban networks and proceed to an empirical analysis of the network used as a case study. We then proceed to a sensitivity analysis and exploration of the model on synthetic data. The model is calibrated on real European firm ownership data. We finally discuss implications of our results and possible extensions.

\section*{Materials and Methods}

\subsection*{Generative urban network model}

\subsubsection*{Rationale}

The model is expected to capture a single macroscopic urban scale, i.e. the level of the urban system where basic entities are cities. Links are induced by underlying firms but these are not explicit. Cities are defined by their economic size, but also by an industrial sector profile. As the authors of \cite{martinus2018global} point out, a combination of several distances may play a role in establishing linkages. Geographical distance, industrial proximity and complementarity will typically be equally significant \cite{cottineau2020nested}. Therefore, we combine several processes driving network growth: (i) geographical proximity (in our case the crow-fly distance, but it can be generalized to any type of accessibility); (ii) geopolitical proximity, which captures the higher propensity to establish links within existing submarkets within the European common market (confirmed relevant by the fixed effect regression described below); (iii) city economic size, which presents agglomeration effects; (iv) industrial proximity in terms of firm sector composition; and (v) previous linkages (influence of already existing links in the past). This last point is crucial to capture accumulated effects and include path-dependencies, justifying further the use of a simulation model instead of a more simple statistical model.

\subsubsection*{Model description}

Let $1 \leq i \leq N$ cities be defined at time $t$ by their economic structure such that $E_i(t)$ is the total economic volume (GDP) for city $i$ and $S_{ik}(t)$ is a matrix giving economic volume proportions within each economic sector $k$, assuming there are $K$ economic sectors. Formally, the model creates links between cities which are characterized by their economic size $E_i$ (GDP) and economic structure $S_{ik}$ in terms of activity sectors (probability distribution of firms within $K$ sectors). Starting with an existing network with no links, the model iteratively adds links, following a probability given by a generalized Cobb-Douglas function \cite{vilcu2011geometric} as 

\begin{equation}
p_{ij} \propto \left(\frac{E_{i}}{E}\right)^{\gamma_O} \cdot \left(\frac{E_{j}}{E}\right)^{\gamma_D} \cdot \left(\frac{w_{ij}}{W}\right)^{\gamma_W} \cdot s\left(S_{ik},S_{jk}\right)^{\gamma_S} \cdot \exp \left(- d_{ij} / d_0\right) \cdot \exp \left(- c_{ij} / c_0\right)
\end{equation}
where $E  =  \sum_k E_k$, $W  = \sum_{i,j} w_{ij}$ weights of previous links, $s(S_{ik},S_{jk})$ is a similarity function between activity sectors $S_{ik}$ and $S_{jk}$ taken as a cosine similarity, $d_{ij}$ is the Euclidian distance, and $c_{ij}$ is a socio-cultural distance. In the case of an already existing link between two areas, the weight of the latter is incremented by a unit volume $w_0$. The model is stopped either when a final time is reached or when the cumulated volume of links reaches a target quantity. This formulation can be considered both as an economic utility model and a generalisation of spatial interaction models.

Note that: (i) we consider an asymmetric influence of sizes, assuming that link directions are important (similarly the similarity function $s$ may be taken as asymmetric); (ii) the influence of previous links is similar to a preferential attachment process; (iii) we do not renormalise the exponents to 1 to ensure to include convex functions. The ``geopolitical/sociocultural'' distance remains abstract and should be parametrised (see setup below) and estimated in the data-driven setting. In this simple version of the model we assume an absence of evolution of economic size, which means that the adjusted parameters are valid on a restricted time period.

\subsubsection*{Model setup}

When setting the model, it is important to consider that different initial conditions (or parametrisation on real data), given by the spatial distribution, the sizes and industrial sector compositions, are going to affect the behaviour of the model. One of the authors \cite{raimbault2019space} has already shown that the initial spatial configuration can have significant impacts on model outcomes. We therefore consider two versions: one with a synthetic system of cities which can be stochastically repeated, and one other based on real data.

\paragraph{Synthetic setup}

The synthetic system of cities is generated to be at the scale of a continent with properties similar to Europe: (i) $N=700$ cities with an empirical power-law from the Global Human Settlements database for economic sizes ($\alpha = 1.1$), distributed randomly into a uniform and isotropic square space of width $d_{max}=3000km$;
(ii) clustered into $C = 20$ countries using a k-means algorithm (this remains consistent with non-correlated city sizes at the scale of the countries as our cities are randomly distributed in space \cite{simini2019testing});
(iii) with synthetic distribution of sectors following log-normal distributions adjusted in a way that large cities have more high-value activities and are more diverse.

More precisely for this last point, assuming sectors are distributed continuously along a one-dimensional axis, cities are assumed to have a sector distribution with a log-normal density function. The largest city is constrained to have a standard deviation and a mode of $1/2$ and the smallest $1/K$, with a linear interpolation between the two for other cities. Parameters of the log-normal law are obtained by solving the corresponding two equations system for each city, and each distribution is discretised and renormalised to be a vector of probabilities. See S2 Text for all details of this synthetic setup.

In synthetic experiments, the network is generated from an initial complete network with dummy links of weight 1 until reaching $t_f=1500$, with a unit link volume of $w_0 = 1$.

\paragraph{Real setup}

The real setup is done using the empirical network of ownership links. Urban areas are distributed according to their positions in the actual geographical space, and economic size and sector composition attributes are used. The socio-economic distance is constructed using information from the statistical modeling, by considering fixed-effect coefficients estimated for each pair of countries. See S2 Text for more details on the construction of the socio-economic distance matrix.

The initial network is the same than in the synthetic setup, and the real links are taken as targets for the generated network. We constrain the total volume of flows by setting $w_0 = \sum_{ij} w_{ij}^{(obs)} / t_{f}$ when the number of time steps $t_f$ is given as a parameter.

\subsection*{Model parameters and indicators}

The model parameters are the Cobb-Douglas weights $\gamma_O,\gamma_D,\gamma_W,\gamma_S$ and distance decay parameters $d_0,c_0$. As the model also includes path-dependencies by integrating the influence of previous links, it cannot be reduced to a simple statistical model. Finally, the role of space introduces even further complexity, yielding a generative model, which has to be simulated.

We consider two main macroscopic indicators to quantify the generated network. Firstly, geographical structures are captured by internationalisation (modularity of countries in the network). The situation of a minimised modularity corresponds to dense connections between cities of different countries in Europe and sparse connections between cities of the same European country. In practice, modularity is computed as follows
\begin{equation}
    I = \frac{1}{2 \left|E\right|} \cdot \sum_{i,j} \left(w_{ij} - \frac{d_i d_j}{2 \left|E\right|}\right) \mathbbm{1}_{c_i = c_j}
\end{equation}
where $d_i$ is the weighted degree of city $i$ in country $c_i$.

Secondly, we consider the metropolisation indicator as the correlation between the weighted degree and the GDP of the city. With this indicator we capture the effect of large cities or metropolises of regional importance attracting the biggest amount of links and involved in high interactions with other cities. It is computed with a Pearson correlation estimator as
\begin{equation}
    M = \hat{\rho} \left[d_i , E_i \right]
\end{equation}

In this work, we do not consider the relationship between countries given their economic complexity, such as the framework developed by  \cite{hidalgo2007product}, where the diversity of products produced plays an essential role in determining co-dependencies and resilience.
What we investigate instead, is the role of linkages given by geography and history, and hence, the sector structure of cities does not evolve within the model.

\subsection*{Empirical data}

Companies from \emph {AMADEUS} are georeferenced using the Geonames database (using postcode or address depending on the information available). We then join this data with the Global Human Settlement dataset GHS-FUA for Functional Urban Areas \cite{Florczyk2019ghs} -- which delineate the commuting areas for urban centres. Starting from a firm-level dataset of 1,562,862 nodes and 1,866,936 links, we obtain a directed network of 573 urban areas and 9323 ownership links. The weight of links is obtained by computing the owned share of turnover at destination, i.e. $w_{ij} = \sum_{k \in i,l \in j} p_{kl} \cdot T_l$ where $p_{kl}$ is the ownership share of company $l$ by company $k$ and $T_l$ is the turnover of company $l$. Node attributes include the economic size of urban areas, that we compute as the sum of turnovers of companies within the area. This quantity is highly correlated with the GDP of the area-level included in the GHS database (Spearman correlation $\rho = 0.71$) and also with population ($\rho = 0.60$). We then expect to capture size effects while keeping the consistence of a single main data source. Another attribute of the nodes is the industrial sector composition, expressed as a proportion of turnover in the area associated to a given industrial sector. Detailed sectors of firms are available up to the four digits of the NACE classification in the raw data \cite{EUROSTAT2008}. We consider the first level classification (21 categories) and compute sector proportions accordingly. This allows us to define an industrial proximity matrix between urban areas, taking a cosine similarity between vectors as $s_{ij} = \sum_{k=1}^{K} S_{ik}\cdot S_{jk}$ where $S_{ik}$ are sector proportions.

\begin{figure}
    \begin{center}
        \includegraphics[width=\linewidth]{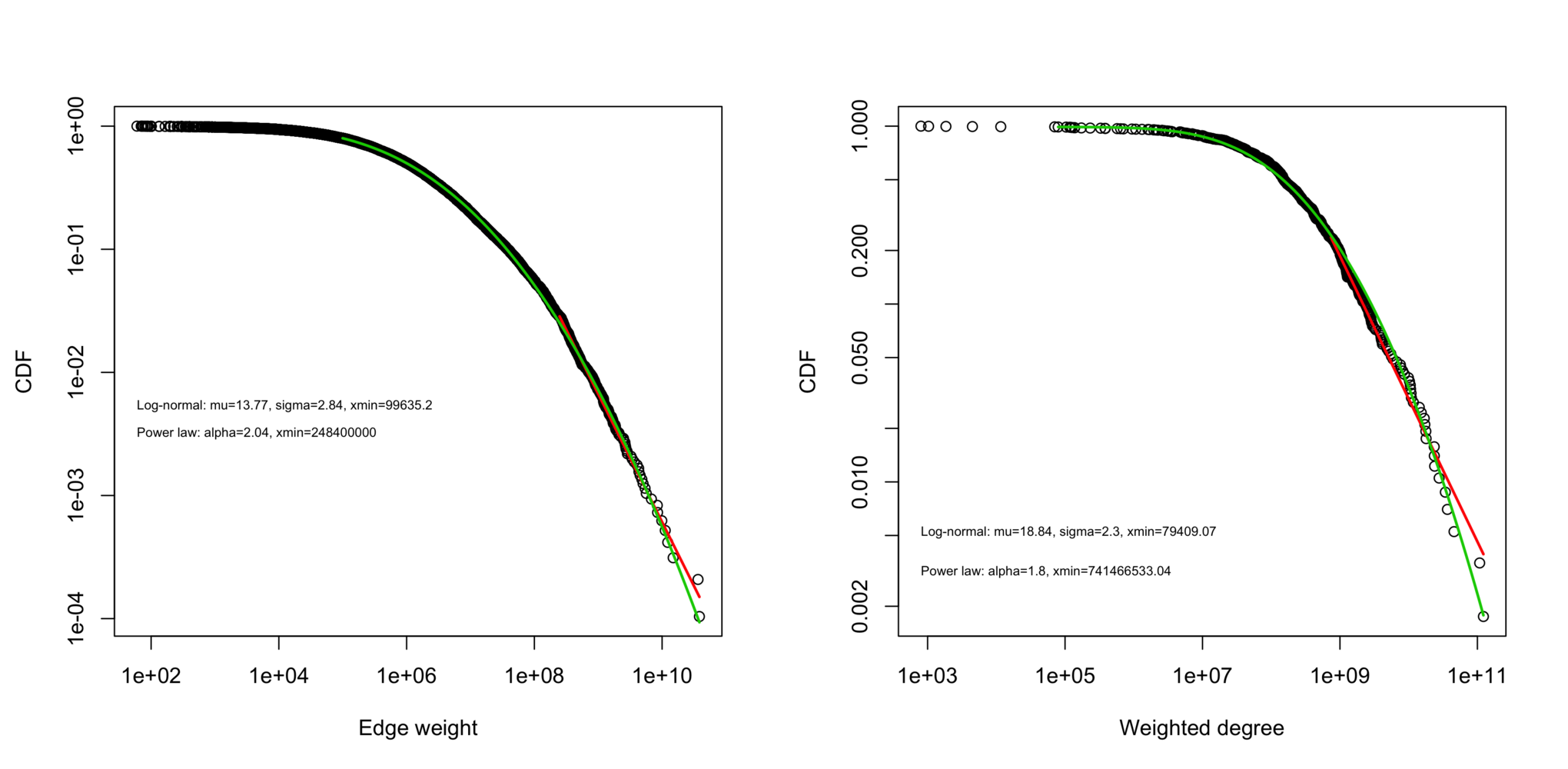}
    \end{center}
    \caption{Distributions of network properties. \textit{(Left)} Cumulated distribution for edge weight, with power law fit (red) and log-normal fit (green); \textit{(Right)} Cumulated distribution for weighted degree.}
    \label{fig:nwdist}
\end{figure}

The urban network has heavy tail weighted degree and edge weight distributions, as shown in Fig.~\ref{fig:nwdist}. We fit power law and log-normal distributions, including minimal value cutoff following \cite{clauset2009power}, using the \texttt{powerlaw} R package \cite{powerlawpackage}. For both, log-normal distributions appear to be a better fit ($\mu=18.8$, $\sigma=2.3$ for weighted degree; $\mu=13.8$, $\sigma=2.8$ for edge weight), and include a much broader range of empirical distributions with lower estimated cut-offs. These heavy tail properties suggest the relevance of a generative model including self-reinforcing processes, which are known to produce such distributions.

%%%%%%%%%%%%%
%%%%%%%%%%%%%

\begin{figure}
    \begin{center}
    \includegraphics[width=\linewidth]{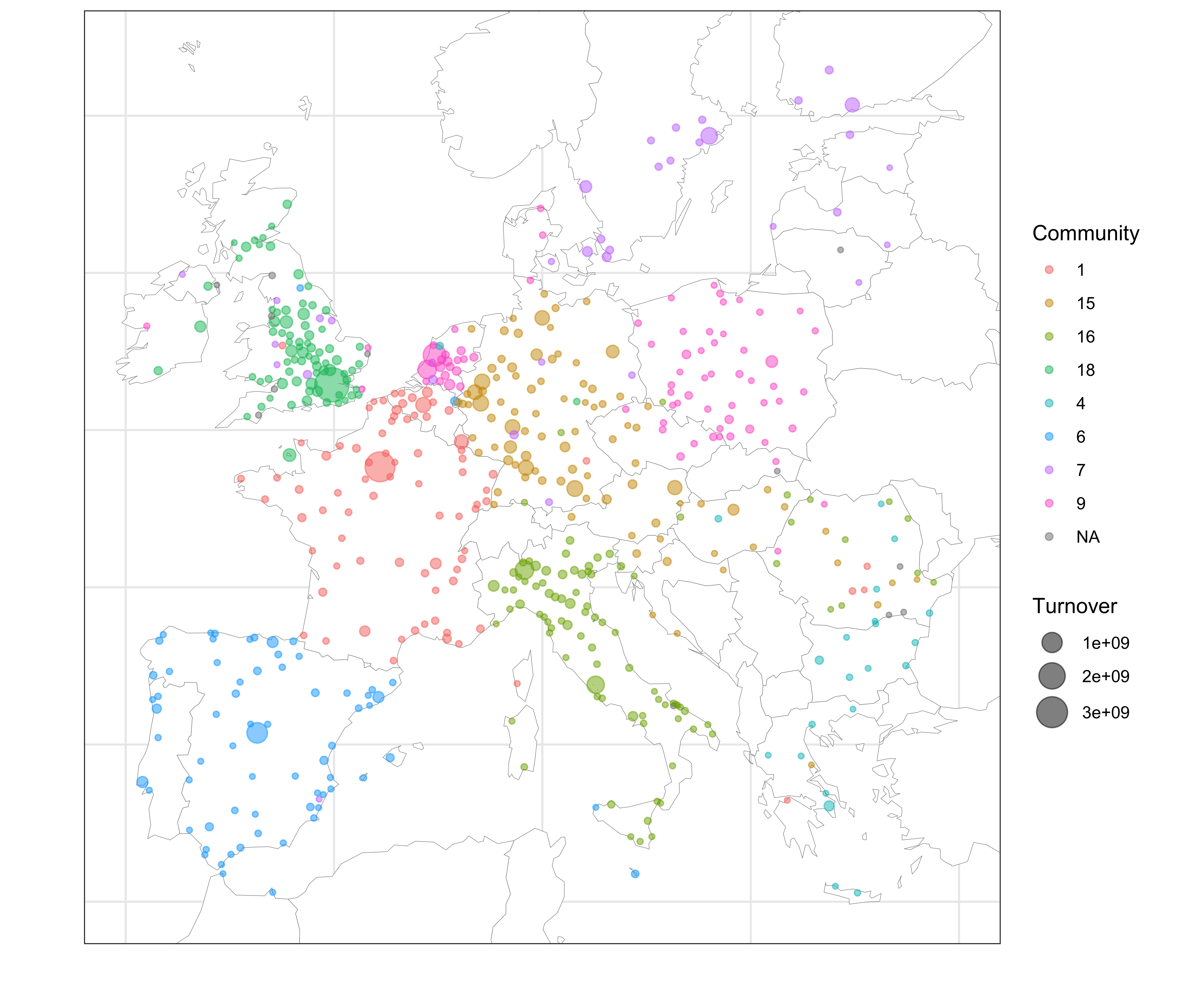}
    \end{center}
    \caption{{\bf Map of network communities obtained by modularity maximisation.} The area of circles gives the turnover within the Functional Urban Area, and color the community it belongs to. Urban Areas in communities of size smaller than 3 were not colored for readability.}
    \label{fig:fig2}
\end{figure}

When focusing on the community structure of the network, different modularity maximisation algorithms (ran on the undirected corresponding network) result in the same undirected modularity of 0.38, a directed modularity \cite{nicosia2009extending} of 0.34 for the greedy algorithm of \cite{clauset2004finding} and of 0.36 for the Louvain algorithm \cite{blondel2008fast}, corresponding to 36 communities (resp. 15). In comparison, when taking into account 29 countries, the classification of urban areas by country has a modularity of 0.32. This indicates that the structure of flows present certain regional patterns. Indeed, a null model assigning random labels with 30 communities, bootstrap 1000 times, yields an average modularity of $0.049 \pm 0.002$, confirming the statistical significance of these communities. We show in Fig.~\ref{fig:fig2} a map of the communities obtained.

The communities reflect characteristic socio-cultural structures and cross-border economic ties throughout Europe. Cities in France, Belgium and Luxembourg belong to the same cluster as they share French as a common business language, but also cross-border tax arrangements creating easier economic interactions \cite{DecovilleDurand2019}. The same socio-cultural proximity effect is seen in cities of the United Kingdom and Ireland, and those of Sweden, Finland, Estonia, Latvia and Lithuania. The cluster composed of Dutch and Polish cities is justified by the fact that firms from the Netherlands are among the first foreign owners of polish firms since 2010 \cite{2019arXiv191014652Z}. The cluster including cities in Germany, Czech Republic, Austria, Slovakia, Hungary and partly in Romania, find its origin in the dependence of the Central and Eastern European economies on German and Austrian large companies mostly from the automotive sector (i.e. Audi, Volkswagen). In addition, all cities of the later cluster have in common to be at the intersection of a very important freight transport route in Europe from Germany to Turkey \cite{Zdanowska2017}. 
%%%%%%%%%%%%%%
%%%%%%%%%%%%%%

\subsubsection*{Statistical analysis}

In order to understand the most determining geo-economic factors influencing the creation of links between cities, we run a statistical analysis as applied in the case of unconstrained spatial interaction models \cite{wilson1975some}. We test five different statistical models, which progressively include the processes for the generative network model. The most general statistical one is:

\begin{equation}
\log(w_{ij}) = \log(d_{ij}) + \log(T_i) + \log(T_j) + \log(s_{ij}) + \alpha_{c_i,c_j} + \varepsilon_{ij}
\end{equation}
where $\alpha_{c_i,c_j}$ is a fixed effect term for the coupling of countries $c_i,c_j$, and $T_i$, $T_j$ correspond to the turnovers of cities $i$ and $j$ respectively. 

This multiplicative form linearly transformed through logarithms is common practice for spatial interaction models. Note that no constraints are considered, since links include information about the turnover and a share of ownership at the same time. 
Estimation results are given in Table~\ref{tab:reg}. We also fit a Poisson generalized linear model with a logarithmic link on the rounded weights \cite{flowerdew1988fitting}. The findings show that when distance only is taken into account (model 1), it has a significant negative effect on link creation, but with an explained variance close to zero. Model 2 adds fixed effects by pair of countries, for which around a third are significant, and with a better $R^2$ of 0.17. Adding the economic size at origin and destination (turnovers $T_i$ and $T$, model 3) slightly improves the explanatory power (model 3). The best model among the linear ones is the model 4, with all factors and fixed effects, both in terms of explained variance ($R^2 = 0.31$) and Akaike information criterion. Model 4 also indicates the absence of overfitting. Model 5, which is a Poisson model with all variables is much a better fit regarding pseudo-$R^2$ and provides much more significant estimates (all fixed effects significant and much smaller standard deviations). Across all models, consistent qualitative stylized facts regarding link creation between cities are revealed: (i) distance has a negative influence; (ii) economic sizes have a positive effect, with the size of the origin city being more important (consistent in the ownership links context, as strongly driven by decision at the headquarter/owner level); (iii) industrial proximity has a positive influence, i.e. similarity is more important than complementarity for this dimension; (iv) a large part of country fixed-effects are statistically significant; and (v) the order of magnitude of the influence of each factor is roughly the same, which means that no process totally dominates the others.
This confirms the relevance of including all these different dimensions in the generative network model, since they have a comparable explanatory role in statistical models.

%%%%%%%%%%%%%
\begin{table}[!ht]
%\begin{adjustwidth}{-1in}{0in}
\caption{{\bf Statistical models.} Ordinary Least Square estimations of statistical models explaining $\log(w_{ij})$ (standard errors in parenthesis). ***, ** and * respectively indicate 0.01, 0.05 and 0.1 levels of significance. For the fixed effect by origin-destination pairs of countries, the proportion of significant coefficients is attributed (where $p<0.1$). Models (1) to (4) are ordinary least square models, while model (5) is a Poisson generalized linear model with logarithmic link. We compare model performance using $R^2$, Mean Square Error, and Akaike Information Criterion. For the Poisson model (model 5), AIC is not given as it is not comparable. \label{tab:reg}}
\medskip
\begin{center}
\begin{tabular}{|l|c|c|c|c|c|}
\hline
Model  & (1) & (2) & (3) & (4) & (5) \\ 
\hline
$\log(d_{ij})$ &      -0.06** (0.03) &   -0.11** (0.05)  & -0.41*** (0.02)  & -0.35*** (0.04)  &  -0.26*** ($5\cdot 10^{-6}$) \\
$\log(T_i)$ &   &   & 0.56*** (0.01) &  0.56*** (0.01) & 0.79*** ($2\cdot 10^{-6}$) \\
$\log(T_j)$ &     &   & 0.39*** (0.01) &  0.39*** (0.01) & 0.66***  ($1.5\cdot 10^{-6}$) \\
$\log(s_{ij})$ &     &   &  &  0.19*** (0.07) & 0.53*** ($9\cdot 10^{-6}$)  \\
Countries &    &  28.3\% &   &  26.5\% & 100\% \\
\hline
$R^2$ &       0.00059   &  0.17 & 0.21 &  0.31  &  0.56 \\
MSE & 7.75 & 6.84 & 6.10 & 5.33 & 8.72 \\
AIC &        44304   &  43578  &  42131  & 41917 &   \\
\hline
\end{tabular}
\end{center}
%\end{adjustwidth}
\end{table}
%%%%%%%%%%%%%

\section*{Model exploration and calibration}

\begin{figure}
    \begin{center}
        \includegraphics[width=\linewidth]{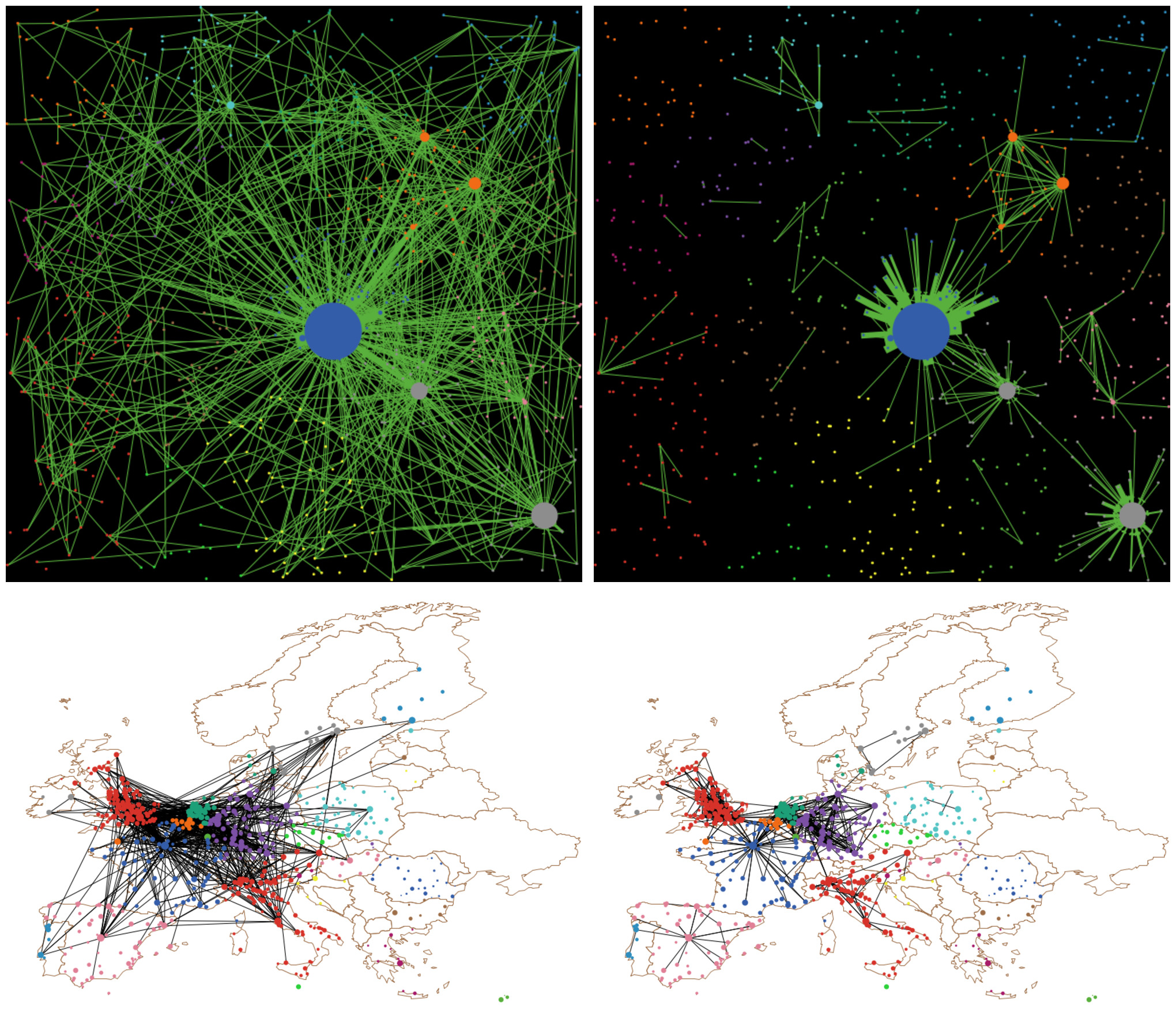}
    \end{center}
    \caption{Example of simulated networks at $t=1500$, for a synthetic system of cities (Top row) and the European urban network (Bottom row), with a high (resp. low) gravity decay parameter for the left column (resp. right).\label{fig:fig3}}
\end{figure}

The model is implemented in NetLogo, which provides a good compromise between performance and interactive prototyping and exploration \cite{railsback2017improving}. Numerical experiments are done by integrating the model into the OpenMOLE software for model exploration \cite{reuillon2013openmole}, which provides sensitivity analysis and calibration methods and a transparent access to high-performance computing infrastructures. The model source code, the aggregated network data (the raw initial firm network cannot be made available for data ownership reasons), and the results are available on the open Git repository of the project at \url{https://github.com/JusteRaimbault/ABMCitiesFirms}. Large simulation result files are available on the dataverse repository at \url{https://doi.org/10.7910/DVN/UPX23S}. We show in Fig.~\ref{fig:fig3} examples of generated networks, with very contrasted patterns obtained with different spatial interaction ranges. Low values of the gravity decay parameter yield local networks around main cities, while long interaction ranges yield more integrated networks. We also show networks simulated in the real setting.

%%%%%%%%%%%%%
%%%%%%%%%%%%%

\subsection*{Synthetic city systems}

We first study extensively the behavior of the model on synthetic systems of cities, to isolate its intrinsic behavior independently of a contingent geographical situation.

\paragraph{Sensitivity analysis}

A global sensitivity analysis first unveils an aggregated influence of parameters on model indicators. We apply Saltelli's Global Sensitivity method introduced by \cite{saltelli2008global}. This produces values for each parameter and each output of a first order sensitivity index, defined as $Var\left[\mathbb{E}_{\mathbf{X}_{\sim i}}\left(Y_j | X_i\right)\right] / Var\left[Y_j\right]$, for parameter $X_i$ and output $Y_j$, $\mathbf{X}_{\sim i}$ being all other parameters. This captures the total influence of $X_i$ all other things being equal. The total order index is given by $\mathbb{E}_{\mathbf{X}_{\sim i}} \left[Var(Y_j | \mathbf{X}_{\sim i})\right] / Var\left[Y_j\right]$ and captures non-homogeneity in behavior and interaction between parameters. Indices were estimated with a Sobol sequence sampling of 20,000 model runs \cite{saltelli2010variance}. We give in S2 Text the full estimated results for sensitivity indices. To summarize, we find that indicators related to modularity (internationalisation and optimal modularity) are mostly influenced by geographical parameters. On the contrary, indicators of network structure are driven by origin and destination sizes. The influence of sectors and historical links has a secondary but not negligible role. Altogether, the analysis witnesses of strong interaction effects between parameters, since for several indicators the total order index is one order of magnitude larger than the first order index. This means that the generative model captures some complex behavior of the system.

\paragraph{One factor sampling}

\begin{figure}
	\begin{center}
	    \includegraphics[width=\linewidth]{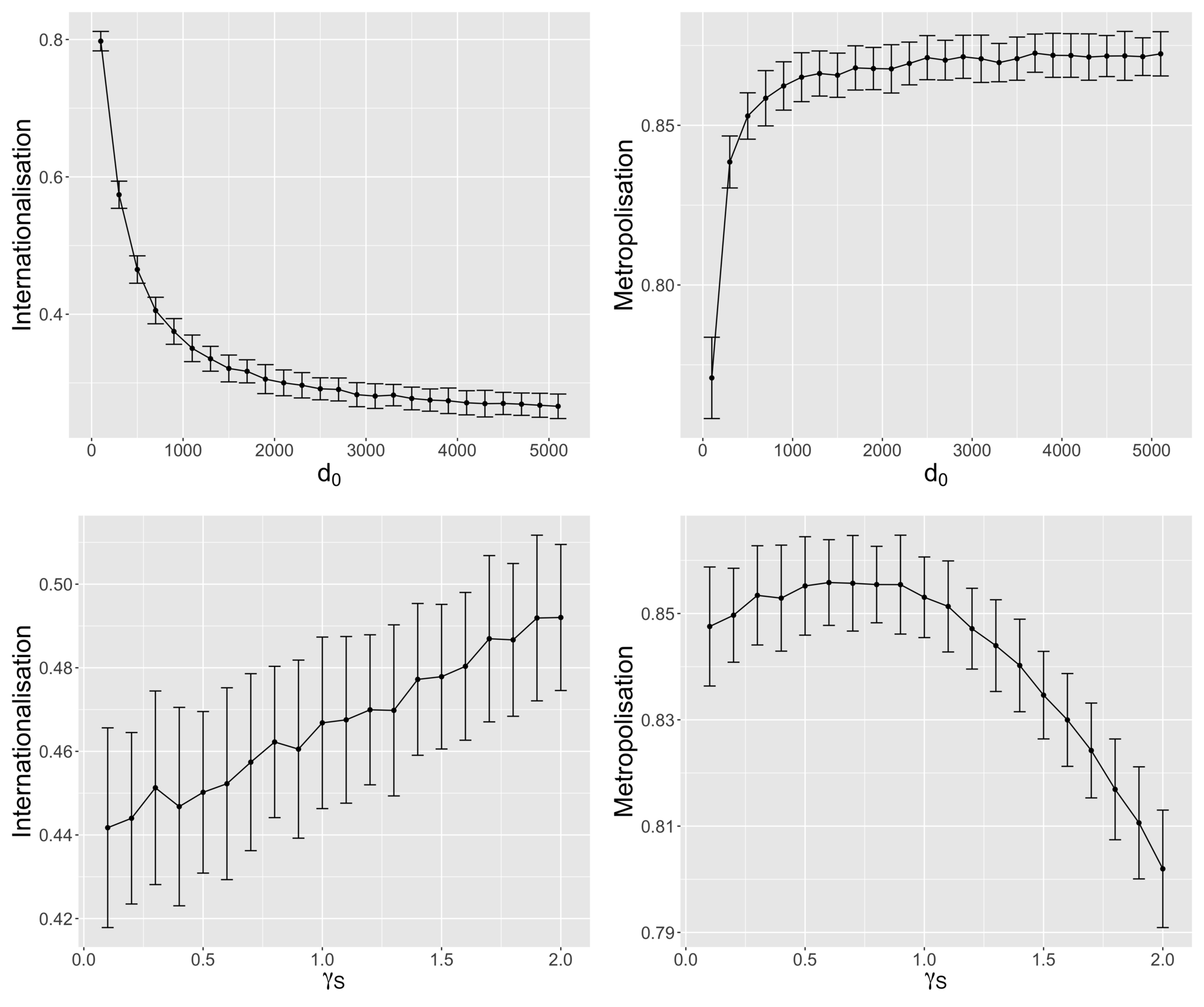}
	\end{center}
    \caption{(Top Left) Internationalisation index decreases exponentially with gravity decay; (Top Right) Correlation between city weighted degree and size. Both plots show a transition from a local to a global regime. (Bottom Left) Internationalisation varies linearly with sector proximity $\gamma_S$; (Bottom Right) Correlation between degree and size exhibits a maximum, witnessing an intermediate regime where size is the most important. \label{fig:fig4}}
\end{figure}

A first numerical experiment of one-factor sampling on all Cobb-Douglas parameters and 100 stochastic repetitions confirms a good statistical convergence (average Sharpe ratios for indicators all larger than 5). We use thus 20 repetitions in following larger experiments (in the case of a normal distribution, a 95\% confidence interval on the average is of size $2\sigma \cdot 1.96 / \sqrt{n}$, what gives $n=16$ repetitions for a CI of width of the standard deviation). Behavior of some indicators are shown in Fig.~\ref{fig:fig4}. For example, internationalisation and metropolisation both show a transition as a function of d$_o$, from a local to a global regime. The sector proximity parameter $\gamma_S$ influences the internationalization linearly, but induces a maximum for the correlation between size and degree, which can be interpreted as a setting where size and the total volume in a city are the most correlated.

%%%%%%%%%%%%%
%%%%%%%%%%%%%

\paragraph{Grid exploration}

\begin{figure}
    \includegraphics[width=\linewidth]{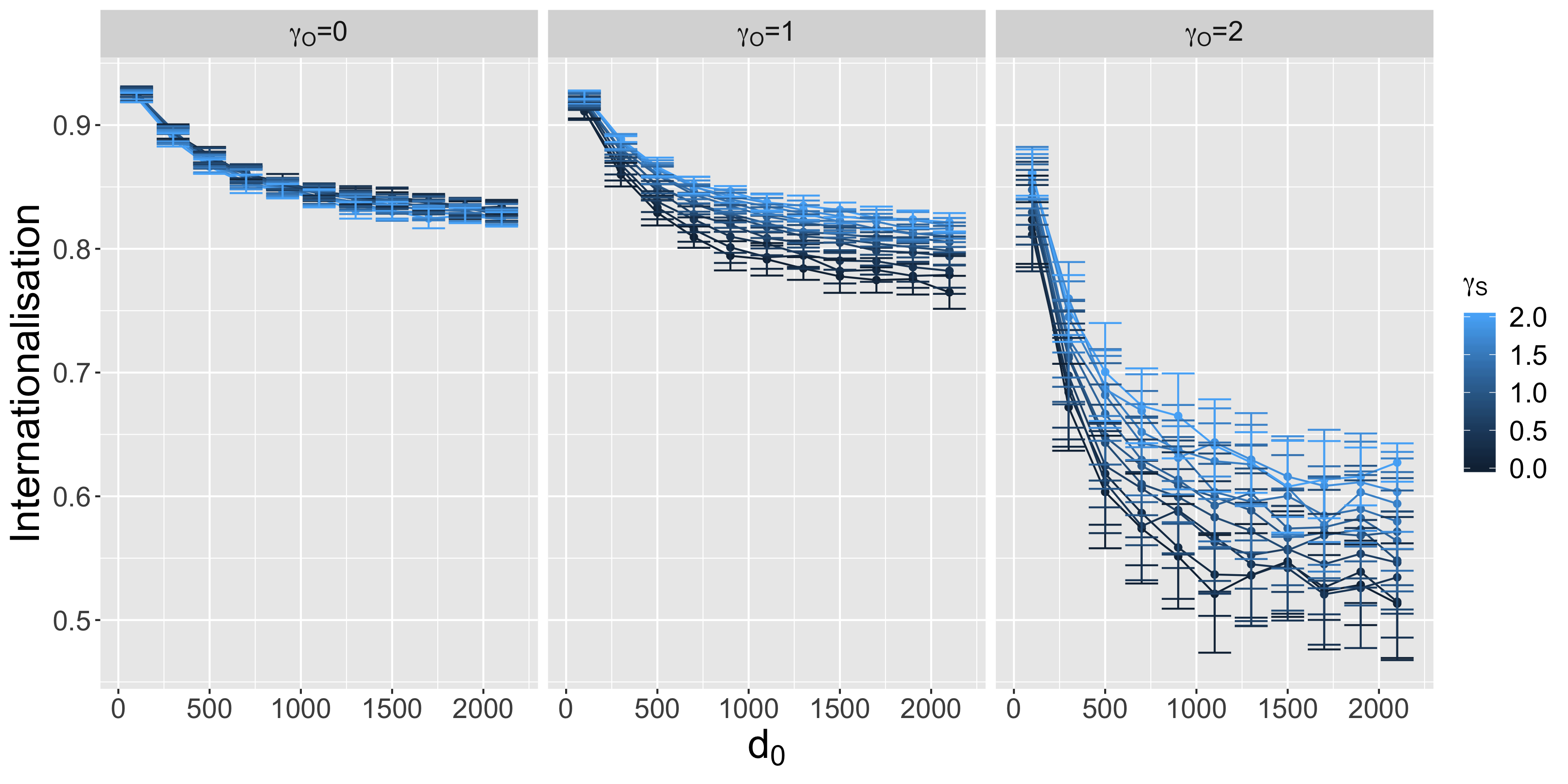}
    \caption{The transition as a function of interaction range depends on the influence of origin size $\gamma_O$; sector proximity $\gamma_S$ plays a role only for a large influence of the origin.\label{fig:fig5}}
\end{figure}

The model behaviour is then studied with a grid sampling for parameters and 20 repetitions (the computations are run on the European Grid Infrastructure, and are equivalent to 2.5 years of CPU time). Distance decay and sector  parameters are varied with 10 steps, other parameters with three. The influence of gravity decay parameters is confirmed when plotting the internationalization index against the distance, which exhibits an exponential decay as shown in Fig.~\ref{fig:fig5}. It interacts with the role of sectors and the size of the origin, with a more significant effect of sector proximity, when size has the largest influence (third panel).

%%%%%%%%%%%%%%
%%%%%%%%%%%%%%

\begin{figure}
    \includegraphics[width=\textwidth]{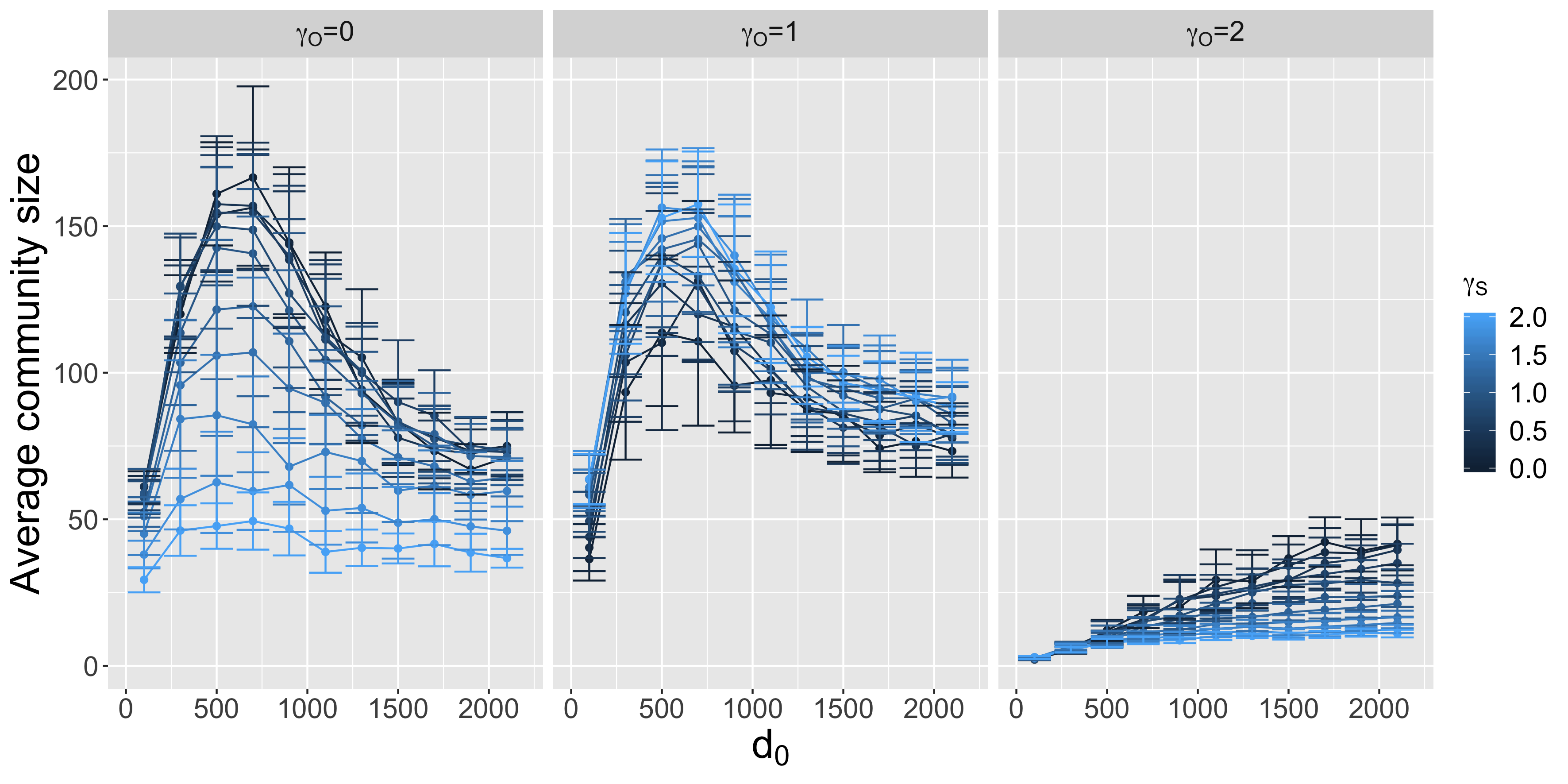}
	\caption{Maximal integration in terms of community size is achieved at an intermediate value of d$_o$: emergence of a regional regime; maximal size depends on the role of sectors $\gamma_S$ in a decreasing way when the origin size is deactivated, and in an increasing way when $\gamma_o=1$; this regime disappears when the influence of the origin size becomes too large.\label{fig:fig6}}
\end{figure}

Other indicators exhibit non-trivial patterns. For example, when considering average community size of the final network as shown in Fig.~\ref{fig:fig6}, we obtain a maximal integration in terms of communities at an intermediate value of the gravity decay. This can be interpreted as the emergence of a regional regime. The size of communities is largely influenced by the value of the elasticities for the similarity function and the ratio of the economic output of the area. In particular, we observe a qualitative inversion of the role of $\gamma_S$ when introducing an effect of the origin (switching $\gamma_O$ from 0 to 1, first panel to second panel). The maximal community size disappears when $\gamma_O = 2$, implying a regime with small local communities where large cities mostly connect with their hinterland. These experiments reveal a complex interplay between processes and how the model produces diverse stylised facts.

%%%%%%%%%%%%%%
%%%%%%%%%%%%%%

\paragraph{Impact of urban hierarchy}

\begin{figure}
    \begin{center}
        \includegraphics[width=\linewidth]{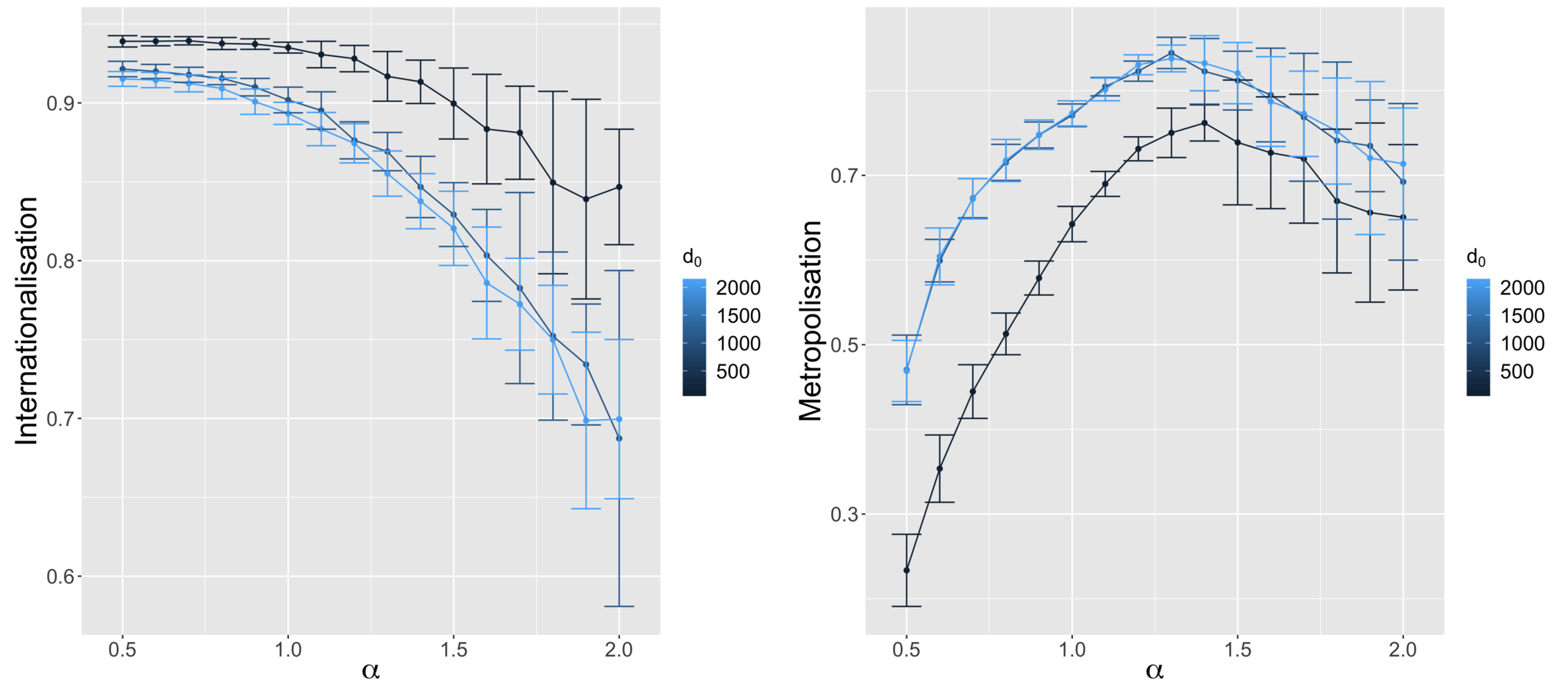}
    \end{center}
    \caption{\textbf{Impact of urban hierarchy.} (Left) Internationalisation as a function of the initial synthetic hierarchy $\alpha$, for different values of interaction range $d_0$ (color); (Right) Metropolisation as a function of $\alpha$, for different values of interaction range $d_0$ (color).\label{fig:fig7}}
\end{figure}

We run a specific experiment to study the impact of urban hierarchy on firms dynamics. Rank-size law in cities is a central stylized fact within urban theories, despite the discussions regarding the estimation of the power-law exponent, its actual values and its variation depending on the definition of cities considered \cite{cottineau2017metazipf,corral2020truncated}. In this particular context, understanding how this parameter can impact urban dynamics within artificial systems of cities -- in which it can be modified to extreme values non observable in real systems--, should bring evidence on underlying processes \cite{raimbault2019space}. We change the initial hierarchy $\alpha$ from 0.5 to 2.0 with a step of 0.1, and also vary the interaction range $d_0$. The behavior of internationalisation and metropolisation indicators are shown in Fig.~\ref{fig:fig7}. As expected, the hierarchy of the urban system has an important impact on model outcomes. We find that more hierarchical urban systems tend to produce more international networks (internationalisation indicator is a modularity which is minimized when networks span more accross different countries), what is consistent with the existence of globalized metropolis \cite{sassen1991global}. This effect is mitigated by lower interaction ranges. Regarding metropolisation, we observe a maximum values as a function of initial hierarchy consistent across different interaction ranges $d_0$. This means that a too high hierarchy is detrimental to the global metropolisation when considering all cities, what may be due to the fact that in such a case very large cities are capturing all flows, leaving nothing to intermediate and medium-sized cities. Interestingly, the optimal value $\alpha \simeq 1.3$ is not far from observed values (which usually range between 0.9 and 1.1 but can take larger values such as 1.5 when extending the urban system \cite{Raimbault_2020}), what may imply that this optimality property is endogenous to the emergence of urban systems and linked to their intrinsic hierarchy.

%%%%%%%%%%%%%
%%%%%%%%%%%%%

\subsection*{Model calibration on the European urban network}

We then apply the model to a real system of cities by calibrating it on the European ownership network. Model setup is done following the real setup described above. The number of time steps $t_f$ is left as an additional parameter, which in a sense determines the granularity of the cumulative network generation process. The objective functions for calibration are the average mean squared error on logarithms of weights given by $\varepsilon_L = \frac{1}{N^2} \sum_{i,j} \left(\log w_{ij} - \log \hat{w}_{ij} \right)^2$ if $\hat{w}_{ij}$ are the simulated weights, and the logarithm of the mean squared error $\varepsilon_M = \log\left(\frac{1}{N^2} \sum_{i,j} \left(w_{ij} - \hat{w}_{ij}\right)^2 \right)$. \cite{raimbault2018indirect} showed that these two are complementary to calibrate urban systems models. Calibration is done using a genetic NSGA2 algorithm \cite{deb2002fast}, which is particularly suited for such a bi-objective calibration of a simulation model. Stochasticity is taken into account by adding the estimation confidence as an additional objective, and we filter the final population to have at least 20 samples. We use a population of 200 individuals and run the algorithm for 50,000 generations, on 1000 islands in parallel.

%%%%%%%%%%%%%%
\begin{figure}
	\begin{center}
    \includegraphics[width=0.75\linewidth]{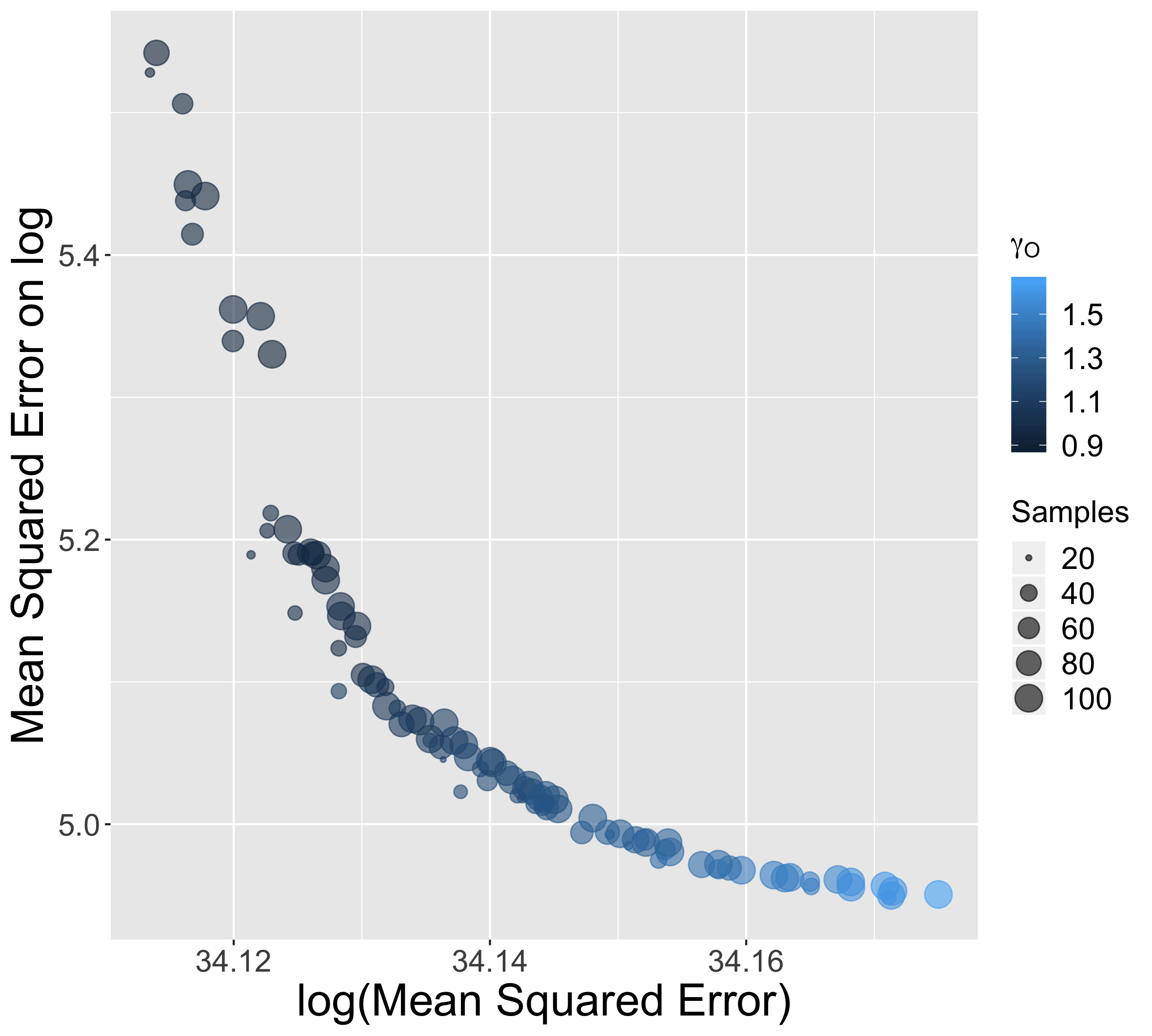}
    \end{center}
	\caption{\textbf{Model calibration results.} Pareto front obtained for the bi-objective model calibration for $\varepsilon_L$ and $\varepsilon_M$ objectives. $\varepsilon_L$ can be compared to the MSE of statistical models fitted above, and the generative model outperform these for several parametrizations (lowest MSE 5.33 for statistical models). Point size gives the number of stochastic samples and color the value of $\gamma_O$.\label{fig:fig8}}
\end{figure}
%%%%%%%%%%%%%%

We show in Fig.~\ref{fig:fig8} the calibration result, as a Pareto front of compromise points between the two objectives. As the mean squared error on logarithm is the same than the one for statistical models, it can be directly compared. The other objective would correspond to statistical models directly on variables: therefore, our model covers in a sense a set of intermediate models which make a compromise between the two objectives, and is therefore much more flexible. The values for $\varepsilon_L$ are for a large proportion of points lower than for statistical models, our model outperforms these in this predictive sense. As shown in Fig.~\ref{fig:fig8}, increasing values for $\gamma_O$ lead to better performance regarding $\varepsilon_L$: increasing the importance of the origin will provide a better fit on all order of magnitudes of link weights (while $\varepsilon_M$ favors mostly the largest links given the fat-tail distribution).

When considering optimal points with $\varepsilon_L < 5.0$, the adjusted parameters are: long range gravity interactions and socio-economic interactions given by $d_0 = 5972 \pm 2148$ and $c_0 = 113 \pm 33$ (the maximal $c_{ij}$ is 11.8); a stronger influence of destination than origin as $\gamma_O = 1.46 \pm 0.10$ and $\gamma_D = 2.0 \pm 0.05$; a comparable influence of sectors as $\gamma_S = 1.76 \pm 0.21$; a fine time granularity as $t_f = 4951 \pm 124$; and a very strong influence of previous links as $\gamma_W = 6.43 \pm 4.11$. This last point furthermore shows in particular a role of the path-dependency parameter, confirming the relevance of this complex generative model including path-dependency. The interaction of this process with others implies an inversion of the relative importance of origin and destination (on the contrary to statistical models). The fact that calibrated interaction distances are large is consistent with long range economic interactions within Europe. Besides these stylized facts, these calibration results altogether demonstrate the feasibility of model application to real data.

\section*{Application: impact of economic shock}

We finally design scenarios to evaluate the impact of economic shocks. These can be the consequence of diverse factors, including socio-economic, geopolitical factors (example of Brexit), or other unpredicted human disasters such as the COVID-19 pandemic. We simulate border closures by rescaling $c_0$ -- the range parameter for socio-economic interactions --, and intra-country lock-downs by rescaling $d_0$ -- the range parameter for direct interactions -- during the simulation, with other parameters being fixed to their default values and the model setup in the real configuration. More precisely, with $t_f = 5000$, the model is run for $t_f / 2$ steps. Then, following two new rate parameters $\kappa_C \in \left[0;1\right]$ and $\kappa_G \in \left[0;1\right]$ defining the shock, model parameters are updated as $d_0 = \kappa_G \cdot d_0$ and $c_0 = \kappa_C \cdot c_0$. The model is then run for the remaining time steps until $t_f$. We study the temporal trajectory of indicators to see the impact of the shock in time. The experiment is run for $(\kappa_G,\kappa_C)$ ranging between 0.2 (strong restrictions) to 1 (reference trajectory), with 100 model repetitions.

\begin{figure}
	\begin{center}
    \includegraphics[width=\linewidth]{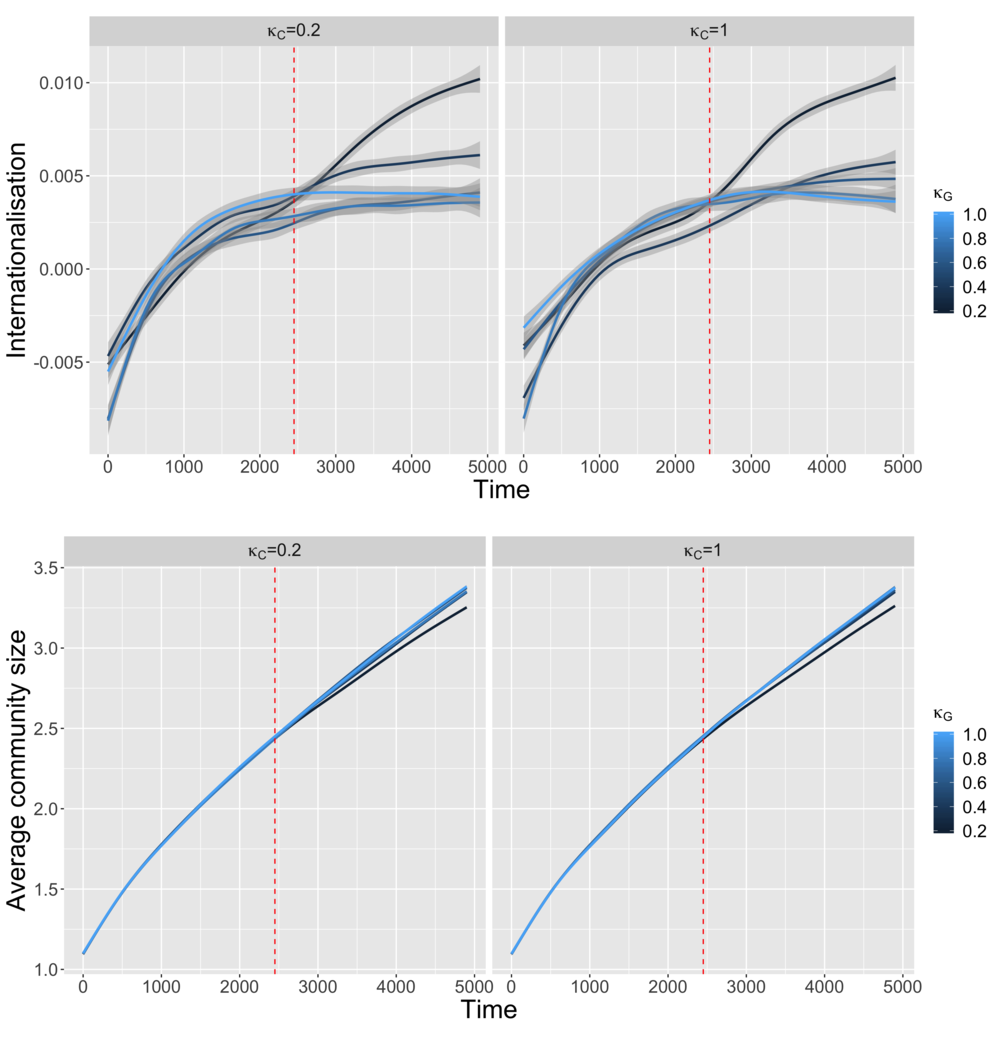}
    \end{center}
	\caption{\textbf{Application of the calibrated model to an economic shock scenario.} All plots show the trajectory of indicators (internationalisation for the top row and average community size for the bottom row) in time, for different values of interaction decay rescaling $\kappa_G$ (color) and socio-economic distance rescaling $\kappa_C$ (panels). The vertical dotted red line shows the moment the economic shock is introduced and $d_0$ (respectively $c_0$) is set to $\kappa_G \cdot d_0$ (resp. $\kappa_C \cdot c_0$).\label{fig:fig9}}
\end{figure}

Results are shown in Fig.~\ref{fig:fig9}. We find an important impact of the shock on internationalisation, but less on the average community size. The impact of changing $d_0$ is more important than $c_0$, but a combination of both yield the stronger effect. For minimal values of $\kappa_G$ and $\kappa_C$, internationalisation is more than doubled, confirming a reconfiguration of urban networks within countries. We validate the statistical significance of effects observed in Fig.~\ref{fig:fig9} by comparing distributions with Kolmogorov-Smirnov (KS) tests. More precisely, we proceed for each indicator and value of $(\kappa_G,\kappa_C)$ to a two-sided KS test between the statistical distribution of the indicator over repetitions at final time with the distribution in the reference configuration. For internationalisation, p-values of the test are all smaller than 0.002 for $\kappa_D < 0.6$ but larger than $0.15$ for $\kappa_D \geq 0.6$. This means that the shock has a statistically significant impact only when interaction range is more than halved. For community size, significance (p-value smaller than 0.01) is reached only for the smallest value $\kappa_D = 0.2$. Thus, the trajectory of the urban network is rather resilient to moderate shocks. This application shows how the model could be applied to practical policy issues.

% computePvalMat(res$internationalizationTS.98)
%             [,1]         [,2]        [,3]         [,4]        [,5]
%[1,] 4.884981e-15 4.440892e-16 0.000000000 0.0000000000 0.000000000
%[2,] 1.364656e-03 9.570235e-06 0.001364656 0.0002468196 0.002318458
%[3,] 8.127483e-01 6.993742e-01 0.999633292 0.4675585921 0.154538055
%[4,] 5.806178e-01 8.127483e-01 0.366726444 0.6993741999 0.906206390
%[5,] 8.127483e-01 2.809600e-01 0.210551633 0.4675585921 1.000000000
% computePvalMat(res$networkAvgCommunitySizeTS.98)
%             [,1]         [,2]         [,3]         [,4]         [,5]
%[1,] 4.365916e-08 4.705150e-06 3.728748e-05 2.267454e-06 1.071070e-06
%[2,] 5.410262e-02 6.993742e-01 6.302223e-03 1.545381e-01 4.675586e-01
%[3,] 9.670685e-01 1.545381e-01 9.996333e-01 9.996333e-01 4.675586e-01
%[4,] 4.675586e-01 5.806178e-01 4.675586e-01 9.670685e-01 6.993742e-01
%[5,] 9.670685e-01 5.806178e-01 4.675586e-01 6.993742e-01 1.000000e+00

%%%%%%%%%%%%%%
%%%%%%%%%%%%%%

%%%%%%%%%%%%%%%
\section*{Discussion}

This paper aimed at presenting a generative model for urban networks defined by interactions between firms based on synthetic data, simulated via the OpenMole platform and calibrated on real data on ownership linkages of firms for Europe from the Amadeus dataset. The simulation on a synthetic system of cities provides a broad knowledge on model behavior in its parameter space. Even in such a simple model (close to directly tractable stationary state) the behaviour is highly non-linear in many dimensions. Our study shows how crucial model exploration is to overcome hidden parameters (deactivated mechanisms or default parameter values). This paper emphasis on the fact that exploration of intrinsic dynamics on synthetic data is of great importance and should be driven before an application on real data so as to put aside the geographical effects from model dynamics.

The calibration of model suggests its potential application to policy issues and can therefore have practical applications in the future. For example the effect of exogenous shocks in the socio-economic structure can be tested as we did in the stylized application example above. Such shocks can for example be the upcoming impact of Brexit on the European market or the consequence of a long-term closure of the European Union members' borders due to further pandemic outbreaks on the single market. In this sense the UK economic sectors mostly dependent on foreign capital and generating the highest turnovers in 2018, that would be the most affected and worth examining in details are: extraction of crude petroleum, manufacture of tobacco products and of basic pharmaceutical goods (in terms of out-going links from the UK) ; activities of head offices and holding companies, sale of cars and manufacture of motor vehicles (in terms of in-going links to the UK). We showed in our experiment that although the system is resilient to moderate shocks, strong restrictions have an important impact on internationalization and thus can be expected to have a detrimental effect on UK economy due to these foreign ownership links (see also S1 Text).

Several possible extensions of the current work can be undertaken in the future as a co-evolution model with an evolution of city sizes or a model with firm agents leading to a multi-scale agent-based model. A more precise study of the role of path dependency can also be undertaken in order to understand the creation of the future links. Other formulations of the model might be taken into consideration as other formulation of the combination of factors or multi-objective optimisation depending on sectors using Pareto fronts.

\section*{Acknowledgments}

Results obtained in this paper were computed on the vo.complex-system.eu virtual organization of the European Grid Infrastructure (http://www.egi.eu). We thank the European Grid Infrastructure and its supporting National Grid Initiatives (France-Grilles in particular) for providing the technical support and infrastructure. We also acknowledge the funding of the EPSRC grant EP/M023583/1.

%\section*{Supporting information}

\section*{Supporting information: S1 Text}
\label{S1Text}

{\bf UK's economic sectors dependent on international links.} Summary of links between UK and the rest of the EU, by economic sector.

Based on \emph {AMADEUS} dataset, the most vulnerable economic sectors in the United Kingdom have been highlighted. The later are the ones due to be mostly affected by eventual future lock-downs of the international economy (international ownership links). 

When considering the UK sectors generating the highest turnover in 2018 mostly dependent on foreign capital they own (out-going links from the UK - see below Table~\ref{tab:uk_outgoing}) these are extraction of crude petroleum, manufacture of tobacco products and of basic pharmaceutical goods. In terms of in-going links to the UK (see below Table~\ref{tab:uk_ingoing}), activities of head offices and holding companies, sale of cars and manufacture of motor vehicles are sectors mostly dependent on foreign ownership coming from abroad.

\FloatBarrier

\begin{table}[!ht]
%\begin{adjustwidth}{-2.25in}{0in}
\caption{Top 10 UK industries depending on foreign ownership expressed in total turnover (out-going links from the UK)\label{tab:uk_outgoing}}
\centering
\medskip
\begin{tabular}{|l|c|c|c|c|}
\hline
Classification  & NACE Industries & Total turnover \\ 
\hline
1 &     Extraction of crude petroleum &   96,458,258,642  \\
\hline
2 & Manufacture of tobacco products &  5,477,272,464  \\
\hline
3 & Manufacture of basic pharmaceutical products &  4,825,881,930 \\
\hline
4 &  Retail sale in non-specialised stores with food, beverages or tobacco predominating  &   3,681,373,374  \\
\hline
5 &  Wireless telecommunications activities &   3,379,297,218  \\
\hline
6 & Event catering activities & 2,871,593,754  \\
\hline
7 &  Other mining and quarrying n.e.c. &  1,590,548,873  \\
\hline
8 & Business and other management consultancy activities &  1,394,373,014 \\
\hline
9 &  Activities of head offices &  1,322,692,365 \\
\hline
10 & Manufacture of air and spacecraft and related machinery &  1,292,065,497 \\
\hline
\end{tabular}
%\end{adjustwidth}
\end{table}

\begin{table}[!ht]
%\begin{adjustwidth}{-2.25in}{0in}
\caption{Top 10 UK industries depending on foreign ownership expressed in total turnover (in-going links to the UK)\label{tab:uk_ingoing}}
\centering
\medskip
\begin{tabular}{|l|c|c|c|c|}
\hline
Classification  & NACE Industries & Total turnover \\ 
\hline
1 &      Activities of head offices &  358,099,870  \\
\hline
2 & Activities of holding companies  & 259,008,123  \\
\hline
3 & Other business support service activities n.e.c. & 183,098,170 \\
\hline
4 &  Security and commodity contracts brokerage  &  72,482,167 \\
\hline
5 &  Manufacture of motor vehicles  &   65,162,757 \\
\hline
6 &  Sale of cars and light motor vehicles  &    61,019,026  \\
\hline
7 &  Other financial service activities, except insurance and pension funding & 47,003,937 \\
\hline
8 &  Other personal service activities n.e.c.  & 37,926,668  \\
\hline
9 &  Wholesale of solid, liquid and gaseous fuels and related products & 36,693,362 \\
\hline
10 &  Other information technology and computer service activities  &  35,668,779 \\
\hline
\end{tabular}
%\end{adjustwidth}
\end{table}

When expressed in total number of links (Table~\ref{tab:uk_outgoinglinks} and Table~\ref{tab:uk_ingoinglinks}), activities mostly dependent on out-going links from the UK are activities of holding companies, of head offices and fund management activities. In terms of in-going links these are business and consultancy activities and renting and operating of own and leased real estate.

\begin{table}[!ht]
%\begin{adjustwidth}{-2.25in}{0in}
\caption{Top 10 UK industries depending on foreign ownership expressed in total links (out-going links from the UK)\label{tab:uk_outgoinglinks}}
\centering
\medskip
\begin{tabular}{|l|c|c|c|c|}
\hline
Classification  & NACE Industries & Total links \\ 
\hline
1 & Activities of holding companies &   686  \\
\hline
2 & Other business support service activities n.e.c.  &  567  \\
\hline
3 & Activities of head offices &  503 \\
\hline
4 &  Fund management activities &   407  \\
\hline
5 &  Business and other management consultancy activities &  398  \\
\hline
6 &  Extraction of crude petroleum &  366  \\
\hline
7 &  Other activities auxiliary to financial services, except insurance and pension funding & 240 \\
\hline
8 &  Development of building projects  & 221 \\
\hline
9 &  Manufacture of basic pharmaceutical products &  212 \\
\hline
10 &  Other information technology and computer service activities  &  181 \\
\hline
\end{tabular}
%\end{adjustwidth}
\end{table}

\begin{table}[!ht]
%\begin{adjustwidth}{-2.25in}{0in}
\caption{Top 10 UK industries depending on foreign ownership expressed in total links (in-going links to the UK)\label{tab:uk_ingoinglinks}}
\centering
\medskip
\begin{tabular}{|l|c|c|c|c|}
\hline
Classification  & NACE Industries & Total links \\ 
\hline
1 &      Other business support service activities n.e.c. &   16,140  \\
\hline
2 & Renting and operating of own or leased real estate  &  6,320  \\
\hline
3 & Business and other management consultancy activities &  6,094  \\
\hline
4 &  Activities of head offices &   5,628  \\
\hline
5 &  Computer consultancy activities &   5,047  \\
\hline
6 &  Activities of holding companies &  4,728  \\
\hline
7 &  Other personal service activities n.e.c. &  3,996 \\
\hline
8 &  Development of building projects &  3,946 \\
\hline
9 &  Other information technology and computer service activities &  3,757 \\
\hline
10 &  Other financial service activities, except insurance and pension funding  &  3,321 \\
\hline
\end{tabular}
%\end{adjustwidth}
\end{table}

\section*{Supporting information: S2 Text}
\label{S2Text}

{\bf Model setup and sensitivity analysis.} Details on the simulation model: summary statistics for the socio-economic distance matrix, procedure for the synthetic model setup and detailed Global Sensitivity Analysis.

\subsection*{Socio-economic distance matrix}

In the real model setup, socio-economic distance between countries $c_{ij}$ is constructed using fixed effects coefficients from the statistical model. In Table~1 in main text, we estimated several statistical models, including some with fixed effects by pairs of countries. The socio-cultural distance between two countries is then taken as the opposite of the fixed effect coefficient for this pair, for the full statistical model. Summary statistics for the 29x29 matrix (28 EU countries and Jersey which is in the database a distinct country from UK but has an important total turnover since many British companies are located there for fiscal advantage reasons) are shown below in Table~\ref{tab:fixedeff}. We observe that distances are rather localized but with large outliers, and an important proportion are not defined (339 out of 841), corresponding to couple of countries having no exchange at all in the dataset.

\begin{table}[h!]
\caption{\textbf{Summary statistics of the socio-cultural distance estimated with a fixed effects model.}\label{tab:fixedeff}}
\begin{center}
\begin{tabular}{ccccccc}
   Min. & 1st Qu. &  Median &  Mean &  3rd Qu. &  Max. &  NA's \\
 -2.327 &  1.440 &  2.451 &  2.538 &  3.447 & 11.797 &  339 
\end{tabular}
\end{center}
\end{table}

\subsection*{Synthetic model setup}

The procedure for the synthetic sector setup is the following:
\begin{itemize}
    \item Sectors distributions follow log-normal densities with most mass in $\left[0;1\right]$
    \item Large cities are more innovative and more diverse. Assuming that sectors are ordered by innovativity (the larger the sector index the larger the innovativity), this assumption is translated by taking a mode and variance of 0.5 for the largest city and a mode and variance of $1/K$ for the smallest, and a linear interpolation between the two. For each city, we define
    \[
    \log(e_i) = \frac{(\log(E_i) - \log(E_{imin}))}{(\log(E_{imax}) - \log(E_{imin}))} * (1/2 - 1/K) + 1/K
    \]
    \item Log-normal parameters $(\mu_i,\sigma_i)$ for each city are then fixed by $\mu_i - \sigma_i^2 = \log(e_i)$ and $-3 \sigma_i^2 - 2 \log(\exp(\sigma_i^2) - 1) = log(e_i)$
    \item $\sigma_i^2$ is the unique positive root of $f(X)=0$ with $f(X) = -3X - 2 \log(\exp(X) - 1) - \log(e_i)$ and $\mu_i = \log(e_i) + \sigma_i^2$.
\end{itemize}

%solve for mu,sigma of the log-normal as a function of gdp
%       we aim at having a log-normal mostly in [0,1] such that
%       (mode,variance) =  1 / K for the smallest log(E_i) and = 1/2 for the largest log(E_i)
%        writing
%        e_i = (log(E_i) - log(E_imin))/(log(E_imax) - log(E_imin)) * (1/2 - 1/K) + 1/K
%        this yields
%         (1)  mu - sigma^2 = log(e_i)
%         (2) -3 sigma^2 - 2 log(exp(sigma^2) - 1) = log(e_i)
%        => sigma^2 is the unique positive root of f(X)=0 with f(X) = -3X - 2 ln(exp(X) - 1) - ln(e_i)

\subsection*{Sensitivity analysis}

The table~\ref{tab:saltelli} gives the full results for the Global Sensitivity Analysis, for all model indicators and free parameters. We give the first order indices and the total indices.

%%%%%%%%%%%%%
\begin{table}[h!]
%\begin{adjustwidth}{-2.25in}{0in}
\caption{Saltelli sensitivity indices, for indicators in rows and parameters in columns. We give for each pair the first order index (F) and the total order index (T).\label{tab:saltelli}}
\hspace{-1cm}\begin{tabular}{|l|c|c|c|c|c|c|c|c|c|c|c|c|}
\hline
 & \multicolumn{2}{|c|}{$\gamma_G$} & \multicolumn{2}{|c|}{$\gamma_D$} & \multicolumn{2}{|c|}{$\gamma_S$} & \multicolumn{2}{|c|}{$\gamma_W$} & \multicolumn{2}{|c|}{$\gamma_O$} & \multicolumn{2}{|c|}{$\gamma_D$} \\
 & F & T & F & T & F & T & F & T & F & T & F & T \\
 \hline
Internationalisation & 0.2 & 0.3 & 0.7 & 0.7 & 0.001 & 0.009 & $4\cdot 10^{-4}$ & 0.007 & 0.03 & 0.04 & 0.02 & 0.04 \\
Metropolisation & 0.02 & 0.1 & 0.02 & 0.2 & 0.002 & 0.1 & 0.001 & 0.09 & 0.2 & 0.6 & 0.3 & 0.6 \\
Modularity & 0.3 & 0.4 & 0.6 & 0.6 & 0.004 & 0.02 & $3\cdot 10^{-4}$ & 0.01 & 0.005 & 0.03 & 0.002 & 0.03 \\
Avg. com. size & 0.008 & 0.09 & 0.01 & 0.1 & 0.002 & 0.07 & 0.003 & 0.04 & 0.3 & 0.6 & 0.4 & 0.6 \\
Degree entropy & 0.006 & 0.02 & 0.003 & 0.02 & 0.006 & 0.03 & 0.008 & 0.02 & 0.5 & 0.5 & 0.5 & 0.5 \\
Weight entropy & 0.04 & 0.1 & 0.03 & 0.1 & 0.008 & 0.08 & 0.01 & 0.07 & 0.4 & 0.5 & 0.4 & 0.5 \\\hline
\end{tabular}
%\end{adjustwidth}
\end{table}
%%%%%%%%%%%%%

\end{document}